# Centimetre-scale micro-transfer printing to enable heterogeneous integration of thin film lithium niobate with silicon photonics

**Margot Niels,[1,*] Tom Vanackere,[1] Tom Vandekerckhove,[1,2] Stijn Poelman,[1] Tom Reep,[1] Günther Roelkens,[1] Maximilien Billet,[1] Bart Kuyken[1]**

[1]*Photonics Research Group, INTEC, Ghent University - imec, 9052 Ghent, Belgium*
[2]*OPERA-Photonique CP 194/5, Université Libre de Bruxelles, 1050 Brussels, Belgium*
** Margot.Niels@UGent.be*

**Abstract:** The integrated photonics CMOS-compatible silicon nitride (SiN) platform is praised for its low propagation loss, but is limited by its lack of active functionalities such as a strong Pockels coefficient and intrinsic $\chi^{(2)}$ nonlinearity. In this paper, we demonstrate the integration of centimetre-long thin-film lithium niobate (TFLN) devices on a SiN platform using the micro-transfer printing (μTP) method. At a wavelength of 1550 nm, propagation losses of approximately 0.9 dB/cm and transition losses of 1.8 dB per facet were measured. Furthermore, the TFLN was integrated into an imbalanced push-pull Mach-Zehnder modulator, achieving a $V_\pi$ of 3.2 V. The electro-optics nature of the observed modulation is confirmed by measuring the device up to 35 GHz, showing that the printing does not affect the high-speed LN properties.

## 1. Introduction

Photonic integrated circuits (PICs) are predominantly designed on complementary metal oxide semiconductor (CMOS)-compatible platforms, such as silicon (Si) or silicon nitride (SiN). This compatibility facilitates seamless integration with electronic components and enables manufacturing in established foundries [1], which give access to mature platforms and hence potential for low-cost production. Although Si-on-insulator (SOI) is widely used, SiN has gained popularity in the past twenty years as a complementary CMOS-compatible platform [2]. Compared to Si, SiN has an even larger transparency window, e.g. allowing it to operate in shorter wavelengths, enabling applications in life sciences [3,4].

Although the SiN platform excels in accommodating passive structures and exhibits low propagation loss [1,5], it lacks efficient active functionalities [6]. For example, gain, enabling light sources, or photodetection, enabling a link to the electrical domain, are not possible on the SiN platform. Further, SiN has a negligible Pockels coefficient or intrinsic $\chi^{(2)}$ nonlinearity [7], which makes having second order nonlinear conversion and fast, efficient modulators nearly impossible. In some works, these limitations are addressed within the platform itself [8], but also other materials can be integrated onto CMOS-compatible platforms [6]. Lithium niobate (LN), known for its large Pockels effect, presents a promising solution to the modulation speed limitation of CMOS-compatible platforms and to their lack of second order non-linearities.

Numerous publications have shown low loss and good modulation properties for integrated LN circuits [9,10]. Recently, with LN-on-insulator (LNOI) epitaxial stacks becoming commercially available, more and more research is being conducted to cointegrate LN on the CMOS-compatible platforms [4,11]. Thereto, several integration techniques have been developed, which include flip-chip bonding [12], where a prefabricated device is directly bonded to the target wafer; and die-to-wafer [10,13–16] or wafer-to-wafer bonding [11,17], where the device material is bonded to the target wafer, after which the substrate is removed and the device is patterned.

A more recent technique is micro-transfer printing (µTP), which combines aspects of flip-chip and die-to-wafer bonding [18]. In µTP, devices are first fabricated on a source wafer and then pick-and-placed onto a target wafer. This back-end integration method allows, in contrast to bonding, for the preparation of numerous devices on a single source wafer, independent of the circuit layout on the target wafer. This, moreover, reduces the material cost of the source wafer since the source can be densely populated. Due to its typical small size, it is also possible to integrate the device into a local recess. Additionally, devices can be inspected and tested before integration (known-good die concept), ensuring quality and reducing contamination risks to the target wafer's processing line. µTP also supports the integration of multiple different devices of various materials, without interfering with each other's process flow or functionality [18].

The potential of micro-transfer printed LN for heterogeneous integration has been demonstrated in previous studies [19], showcasing the creation of resonators [20–23], second harmonic generation [24] and high-speed modulation on SiN platforms [25]. A summary of micro-transfer printed designs is presented in Table 1. Nonetheless, there remains significant potential to extend the interaction length of these devices, allowing for longer light-matter interaction. Elongated light-matter interaction leads to higher nonlinear conversion efficiency and lower required modulation voltages, and consequently lower power consumption. This reduction in power consumption could further enhance CMOS-compatibility, simplifying integration with existing electronics and advancing the development of high-performance, energy-efficient photonic devices. However, the fabrication and printing of long devices comes with extra challenges. By now demonstrating the successful µTP of cm-scale thin film LN (TFLN), the process is again one step closer to complete CMOS-compatibility. Also, the electro-optic (EO) effect is shown by fabricating some push-pull modulators with a centimetre long interaction length, exhibiting a $V_{\pi}L$ of 3.2 V.cm.

Table 1: Comparison between the reported integration of LN using micro-transfer printing in the literature

| Ref | Tether material | Platform | Size of coupons | Pre-processing of target | Demonstrator | Efficiency |
|---|---|---|---|---|---|---|
| [19] | LN | LN/SiN | 1 mm x 40 µm<br>1 mm x 120 µm | Etching recess | / | / |
| [24] | LN | LN/SiN | 1 mm x 40 µm | Etching recess | PPLN SHG | 2500 %/$Wcm^2$ |
| [20] | Resist | LN/SiN | 100 µm x 100 µm | None | Ring resonator | Q = 32 000 |
| [21] | Resist | LN/SiN | 1.2 mm x 70 µm | None | Micro-cavity | Q = 50 000 |
| [22] | LN | LN/SiN | 110 µm x 60 µm | None | Ring resonator | Q = 100 000 |
| [23] | Resist | LN/Si | 700 µm x 50 µm | None | Ring modulator | Q = 11 000 |
| [25] | LN | LN/SiN | 2 mm x 30 µm | Etching recess | MZI Modulator | $V_{\pi}$=14.8 V |
| [26] | LN | LN/Si | 6.5 mm x 100 µm | None | MZI Modulator | $V_{\pi}$=5 V |
| This work | Resist | LN/SiN | 1 cm x 30 µm | None | MZI Modulator | $V_{\pi}$=3.2 V |

## 2. Fabrication of the source material

The process under consideration has two primary objectives: transferring centimetre-scale slabs of LN and minimising the processing on the target (or destination) wafer. Thereto, the first task at hand is to fabricate a source sample with suspended devices that fits these criteria. Looking at previous work, different approaches can be taken to keep the thin material suspended: tethers to the side [19–21,27] versus tethers to the substrate [21] and LN tethers [19,27] versus resist tethers [20,21]. Also pillars can be added [19] for a more robust release of large devices.

In this work, a process is developed to suspend long devices using resist tethers to the substrate, without the use of pillars. By avoiding the use of pillars, at least one dimension of the device is limited to being relatively small (under 100 μm), but it does not require extra processing on the target platform prior to printing to etch trenches. The use of resist tethers, on the other hand, increases the likelihood that the process can be adopted in semiconductor fabs, whereas LN tethers (and the associated LN breaking) are unlikely to be accepted, except in LN-specific fabs. To validate this process, rectangular thin LN slabs (referred to as coupons) were produced on an LNOI chip (LN/SiOx/Si, Fig. 1.a) and were subsequently transfer-printed onto a patterned SiN chip.

The TFLN device fabrication begins with the etching of the LN layer of an LNOI sample (Fig. 1.a). To that end, an amorphous silicon (aSi) hard mask is applied, patterned via UV lithography, and etched using reactive-ion etching (RIE). Next, the LN is etched through Argon milling and the mask is removed, which results in a situation as shown in Fig. 1.b. Subsequently, the underlying oxide layer is patterned with UV lithography, followed by a dry etch (Fig. 1.c). A resist encapsulation is then applied through spin coating and is patterned through photo-lithography to suspend and protect the coupon during the printing process (Fig. 1.d). In the final step of device creation, the oxide is removed with an HF-based wet etch (Fig. 1.e), leaving the TFLN coupons suspended.

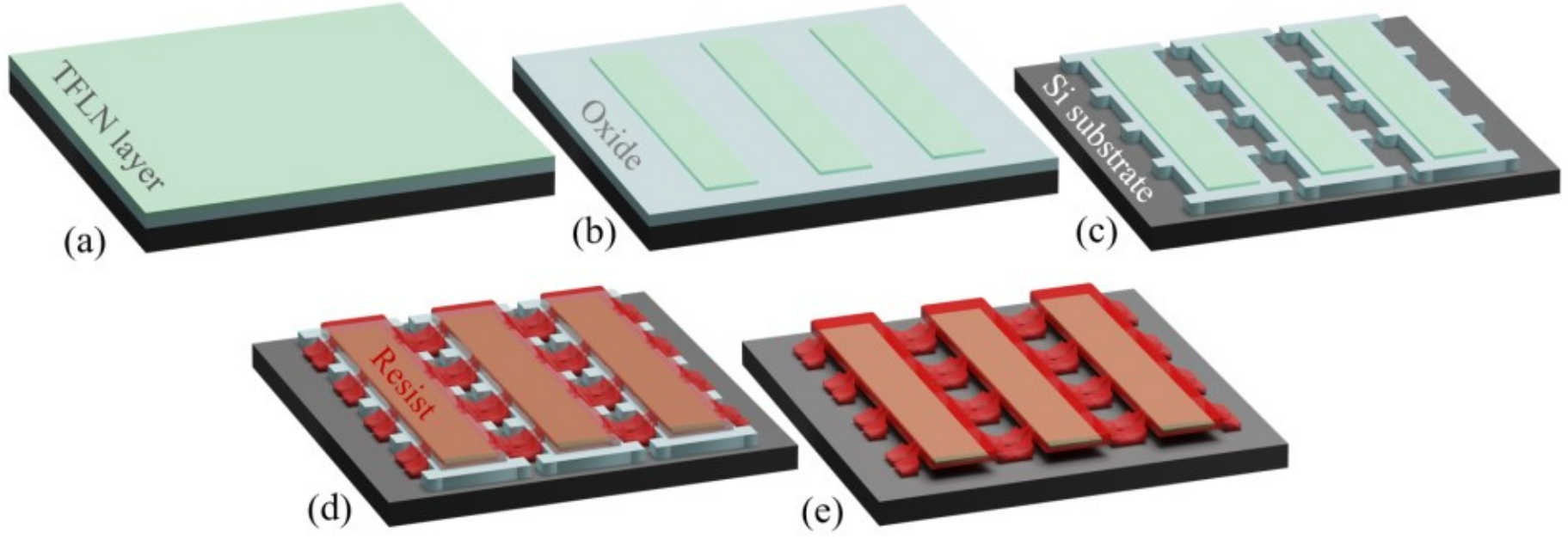

Fig. 1. Process flow for coupon fabrication. (a) LNOI sample (TFLN-on-insulator). (b) The TFLN is patterned. (b) The oxide release layer is etched to the substrate. (d) The devices are encapsulated in photoresist such that they can be suspended. (e) Final suspended TFLN coupons after oxide undercut.

The fabricated coupons can now be transferred to a target substrate. The full printing process is illustrated in Fig. 2: the suspended coupon to be printed is first approached by a Polydimethylsiloxane (PDMS) stamp (Fig. 2.a). The coupon is then picked up by quickly (500 mm/s) retracting the stamp (Fig. 2.b). Next, the coupon is transferred to the target substrate, by pressing it against the substrate and slowly (4 μm/s) retracting the PDMS stamp (Fig. 2.c). Finally, the resist encapsulation is removed to reveal the TFLN (Fig. 2.d) by oxygen plasma.

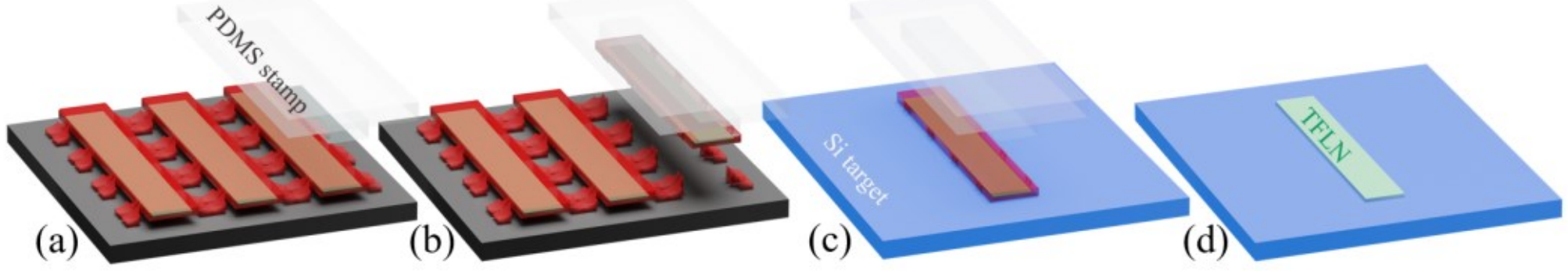

Fig. 2. Printing process for transfer printing of TFLN on a blank Si target. (a) The PDMS stamp approaches the suspended coupon. (b) The coupon is picked up. (c) The coupon is transferred to the Si target. (d) The photoresist encapsulation is removed.

The aforementioned process flow for fabricating the TFLN coupons was put to practice by fabricating a sweep of coupons from 1 mm to 1 cm. An overview of a source sample of suspended TFLN coupons (step Fig. 1.e.) is depicted in Fig. 3.a. and Fig. 3.b shows a close-up view of the coupons, demonstrating a low defect density for a densely populated source sample.

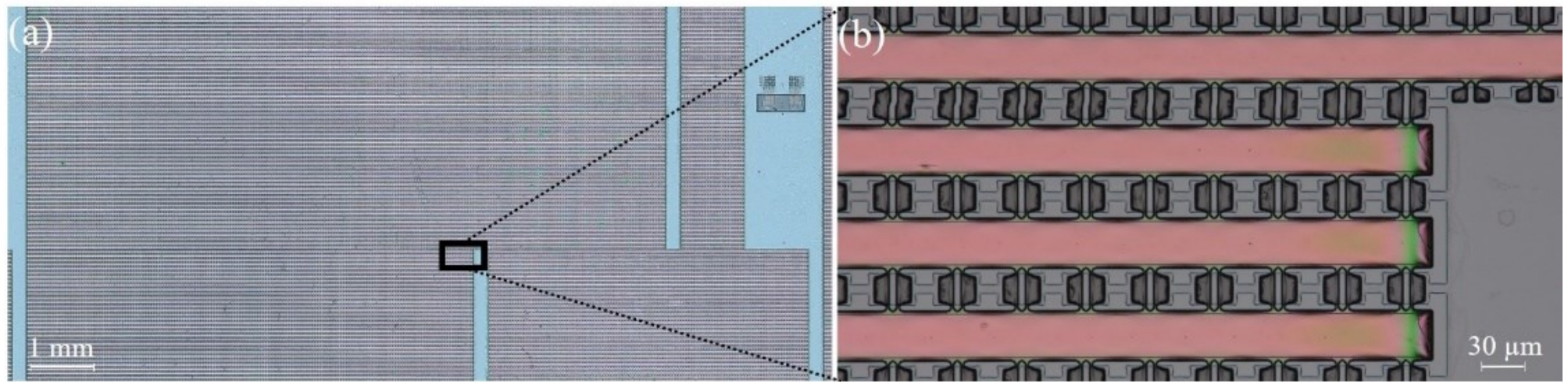

Fig. 3. (a) Overview of a source sample with suspended TFLN coupons. (b) Close-up view of the coupons.

## 3. Loss measurements

The fabricated coupons are printed on patterned targets. To also evaluate the photonic properties, two tests are prepared: a cut-back test to assess the propagation loss contributed by the TFLN coupons and a test involving coupon chains of varying numbers of coupons to evaluate the coupling loss of the coupons. Both tests are conducted on a SiN target sample. A hybrid mode is used where the SiN guides 32% of the light, while 62% of the light is in the X-cut LN (Fig. 4.c). This approach leverages the benefits of LN without requiring etching of LN waveguides, which requires smooth sidewalls in the LN.

For the cut-back structures, the 1.5 µm wide and 300 nm thick SiN waveguides all have the same length, but a different coverage by TFLN coupons. These waveguides are patterned by electron beam lithography and formed by performing an RIE etch.

The printed TFLN coupons vary in length: 1 mm, 2 mm, 5 mm, 7 mm, and 1 cm. They are all 30 µm wide, 300 nm thick and have angled facets to minimise reflections. Since only the length of the LN varies, the propagation losses attributed to the LN can be derived from transmission measurements. Fig. 4 exhibits a close-up of the printing results before (Fig. 4.a) and after (Fig. 4.b) the encapsulation removal, as well as an overview of the printed coupons for a full set of cut-back structures (Fig. 4.d).

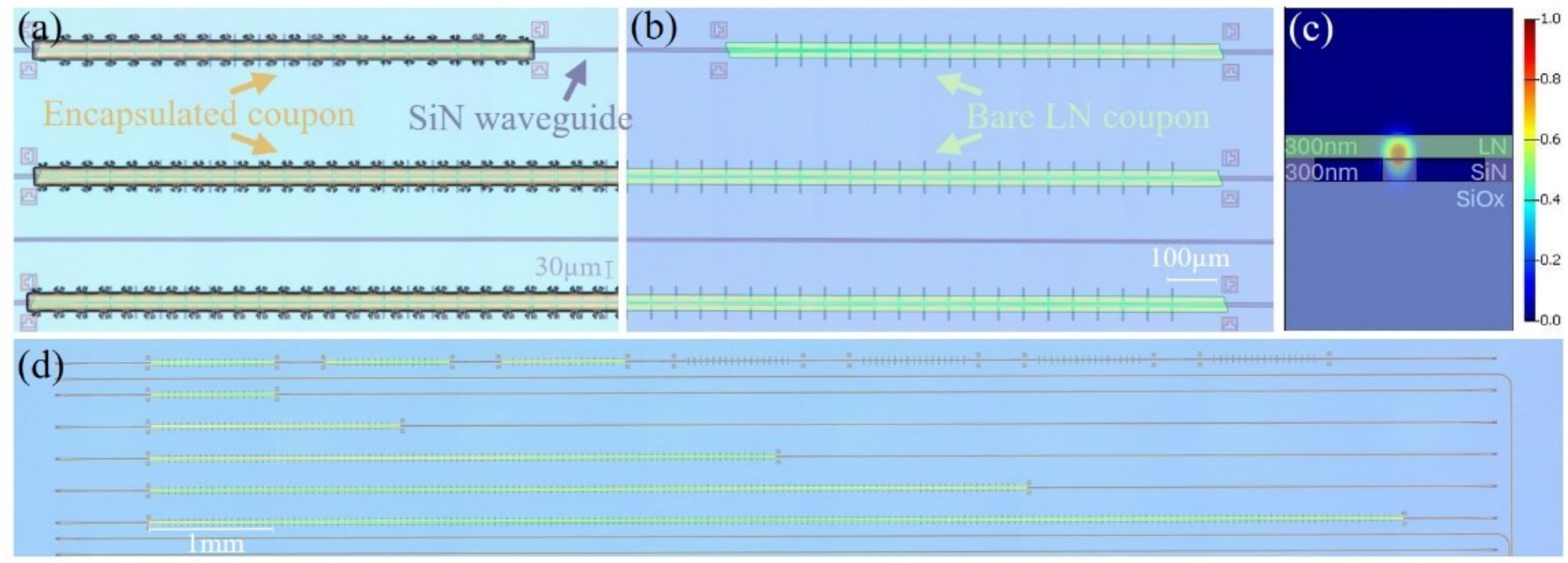

Fig. 4. Printing result for a target with cut-back structures: close-up of printed coupons. (a) Before and (b) after encapsulation removal. (c) Illustration of the hybrid mode. (d) Overview of a full set of test structures.

The chains with the printed coupons, one of which is shown at the top of Fig. 4.d, consist of several TFLN coupons, each measuring 1 mm x 30 µm, linked in series. In this case, three arrangements were printed: no coupon, three coupons (Fig. 4.d) and seven coupons. The wavelength sweeps of all three configurations are plotted together in Fig. 5.c. From these measurements, the reference loss (i.e. propagation loss through the SiN combined with the in-coupling loss of the grating couplers), which is present in each structure, can be determined to be 16.9 dB at 1550 nm. This reference loss is subtracted from the measurement data to normalise the transmission. Comparing than all different configurations at a wavelength of 1550nm, the facet coupling losses are 1.8 ± 0.2 dB per facet on average, as displayed in Fig. 5.a.

Two sets of cut-back structures were populated to account for possible fabrication errors. After photoresist removal, the chip was measured at a C-band wavelength of 1550 nm. Fig. 4.c illustrates the mode profile in the heterogeneous region during the measurements. Excluding one damaged structure (with visible cracks in the LN coupon) from the measurement data, an average propagation loss of 0.9 ± 0.8 dB/cm was found, as shown in Fig. 5.b. The data points in this figure are the average value at 1550 nm for each coupon length. Then a linear fit is added and the slope represents the propagation loss per mm.

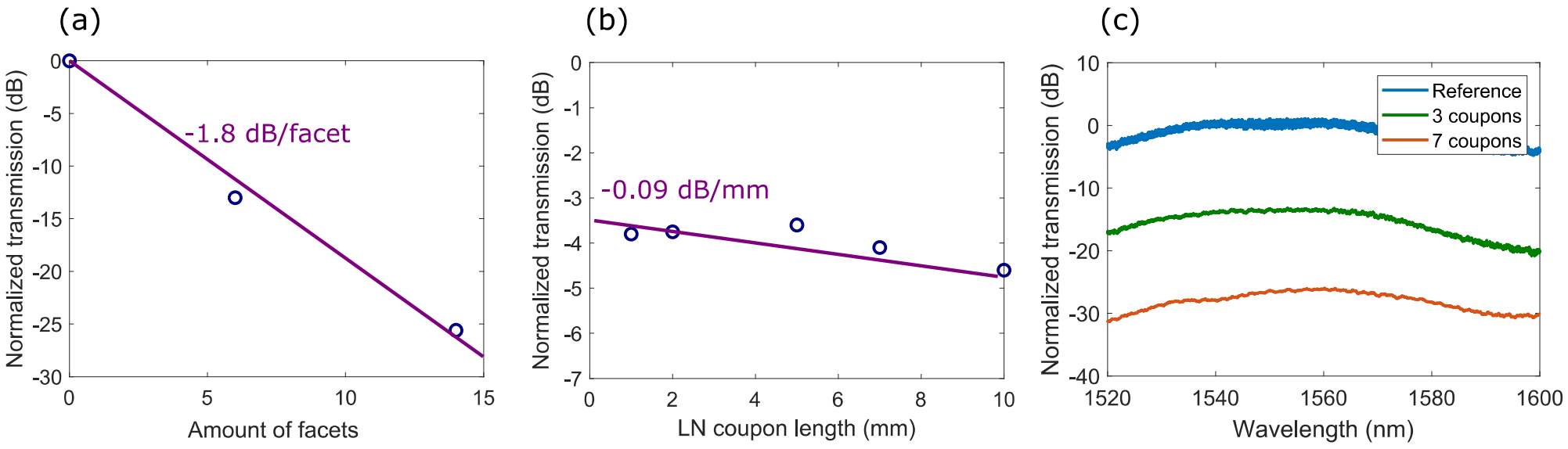

Fig. 5. (a) Transition losses due to abrupt transition between SiN and hybrid SiN-LN. (b) Measurement data for propagation loss in the LN. (c) Wavelength sweep of the coupon chains used to estimate the transition losses.

## 4. Centimetre long modulators

As a demonstration of the technology, an unbalanced Mach-Zehnder modulator (MZM) is designed that contains the fabricated coupons. This design is a push-pull modulator, comprising two phase modulators linked together in parallel by 1x2 multimode interferometers (MMIs). To the best of our knowledge, for the first time, the active electro-optic part of the device, that is based on micro-transfer printing, has a length of 1 cm.

The light is coupled into the 1.5 µm wide and 300 nm high SiN waveguide using grating couplers, which are designed to excite only the fundamental TE-polarised mode around the C-band communication wavelength of 1550 nm. In the modulation sections, this SiN waveguide is covered by a 30 µm wide and 300 nm thick slab of micro-transfer printed X-cut LN. Similar to the approach in the loss structures (Fig. 4.c), a heterogeneous mode is used where the light is guided by the SiN, but is also present in the LN, which adds electro-optic functionalities (Fig. 7.a).

Fig. 6 shows the main processing steps to create the modulators. First, the SiN sample is fabricated, and LN coupons are printed in the designated locations (Fig. 6.a). Next, metal contacts are deposited on the LN coupons, consisting of a 15 nm Ti adhesive layer and a 1 µm Au layer. The gap between the two metal contacts is 6 µm to avoid any transmission loss. A close-up of the coupon ends before (Fig. 6.b) and after metallisation (Fig. 6.c) demonstrates a clean bonding and a successful lift-off process.

The optical characterisation of the MZM begins with a wavelength sweep while no electrical signal is applied to the metal contacts. The result of this sweep is shown in Fig. 7.c, from which an extinction ratio (ER) of at least 31 dB can be extracted. In order to characterise the electro-optic behaviour of the modulator, low-speed measurements are performed. The measurement setup is shown in Fig. 7.e. With the laser set to the quadrature point, 100 Hz sine waves with different peak-to-peak voltages (Vpp) are applied to the metal contacts of both coupons, with one having a phase shift of 180° relative to the other. The transmission is measured by a power meter (PM), from which the output is monitored by an oscilloscope (OSC). Also, the input signal, applied to the metal contacts, from the signal generator (SG) is sent to the oscilloscope.

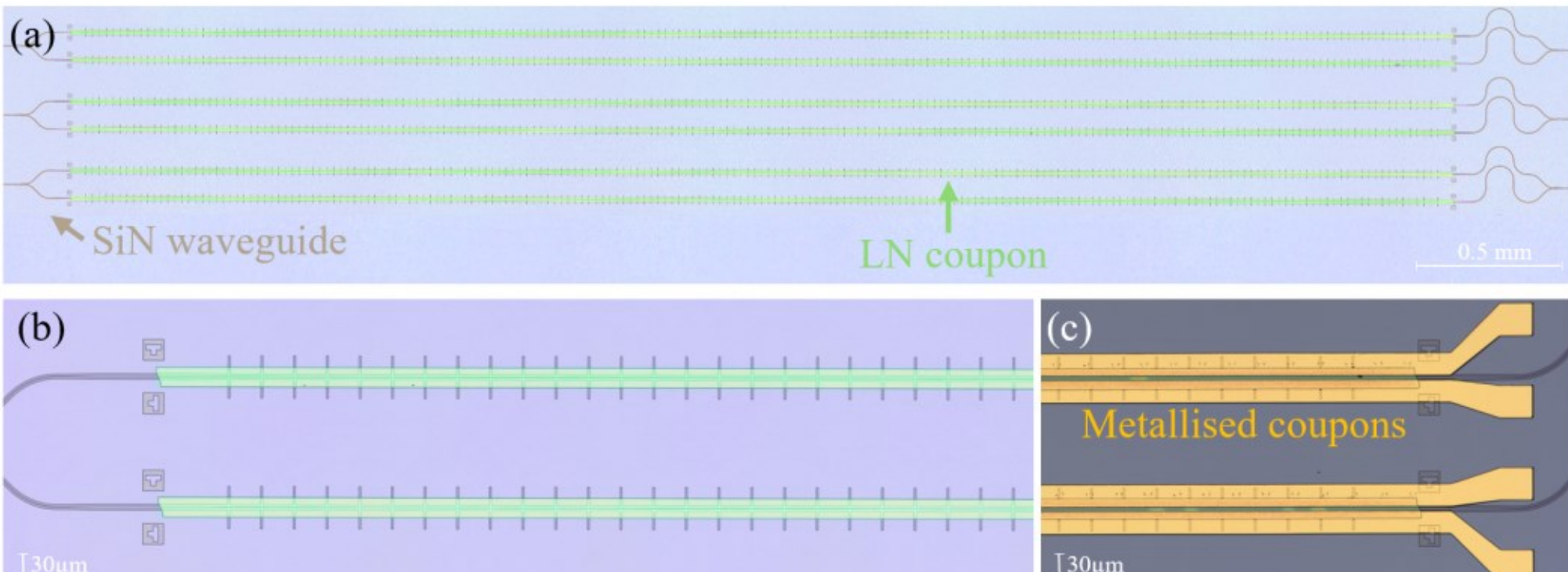

Fig. 6. Printing and post-processing result for a target with modulators. (a) Printed coupons after encapsulation removal. (b) Close-up view of printed the coupons on the SiN circuit. (c) Back-end metallisation, forming the electrodes of the modulator.

First, a sweep of voltages is applied to the metal contacts. This test shows a disformed output as soon as the modulator is overdriven, which provides a first indication of the half-wave voltage ($V_\pi$). This $V_\pi$ value indicates, in turn, the voltage needed to shift the phase in one modulator arm by π. For a better overview, only some of the results are plotted in Fig. 7.b, showing that the modulator is clearly overdriven for an input of 5 Vpp and 7 Vpp. However, the overdrive becomes already visible between 3 Vpp and 4 Vpp. Therefore, to extract the $V_\pi$ value more precisely, an input of 4 Vpp is applied. In Fig. 7.d, both inputs to the oscilloscope are merged, and the normalised transmission is shown as a function of the input voltage. From this graph, a $V_\pi$ value of 3.2 V can be deduced, corresponding to a $V_\pi$ of 6.4 V for a single phase modulator. As this is the result for a 1 cm long push-pull modulator, the $V_\pi$ is the same as the half-wave voltage length product ($V_\pi L$), which is 3.2 V.cm for the push-pull modulator. This value corresponds to the simulated value of 3.2 V.cm following the method presented in [16]. For this calculation, a Γ of 0.42 was used, which is defined as the overlap between the optical and electrical mode. The agreement between measurements and simulation shows the EO nature of the modulation reported here. To the best of our knowledge, this is the lowest reported $V_\pi$ showcased by micro-transfer printing so far.

Compared to the other heterogeneous integration techniques, LN-on-SiN push pull modulators showing a $V_\pi L$ of 2.9 V.cm have been demonstrated using die-to wafer bonding [15]. This is slightly more efficient, but still in the same order of magnitude, while using a less versatile technology. By here utilising μTP as an integration technique for TFLN making low $V_\pi$ devices, the way is paved for new possibilities as co-integration with other active materials. Moreover, we believe that the $V_\pi L$ can be decreased by pushing the boundaries of the electrode gap or increasing the fraction of the mode which is in the LN with respect to the fraction which is in the SiN.

The modulator is then measured at higher speed using a vector network analyser (VNA). To de-embed the device under test (DUT), two measurements are conducted. First, the electro-optic-electric $|S_{21}|$ values (in dB) of the DUT are recorded using a setup as shown in Fig. 8.a. Next, the DUT and the probes are replaced by a commercial modulator (in this case a Fujitsu FTM7937EZ), allowing to extract the contribution of the setup.

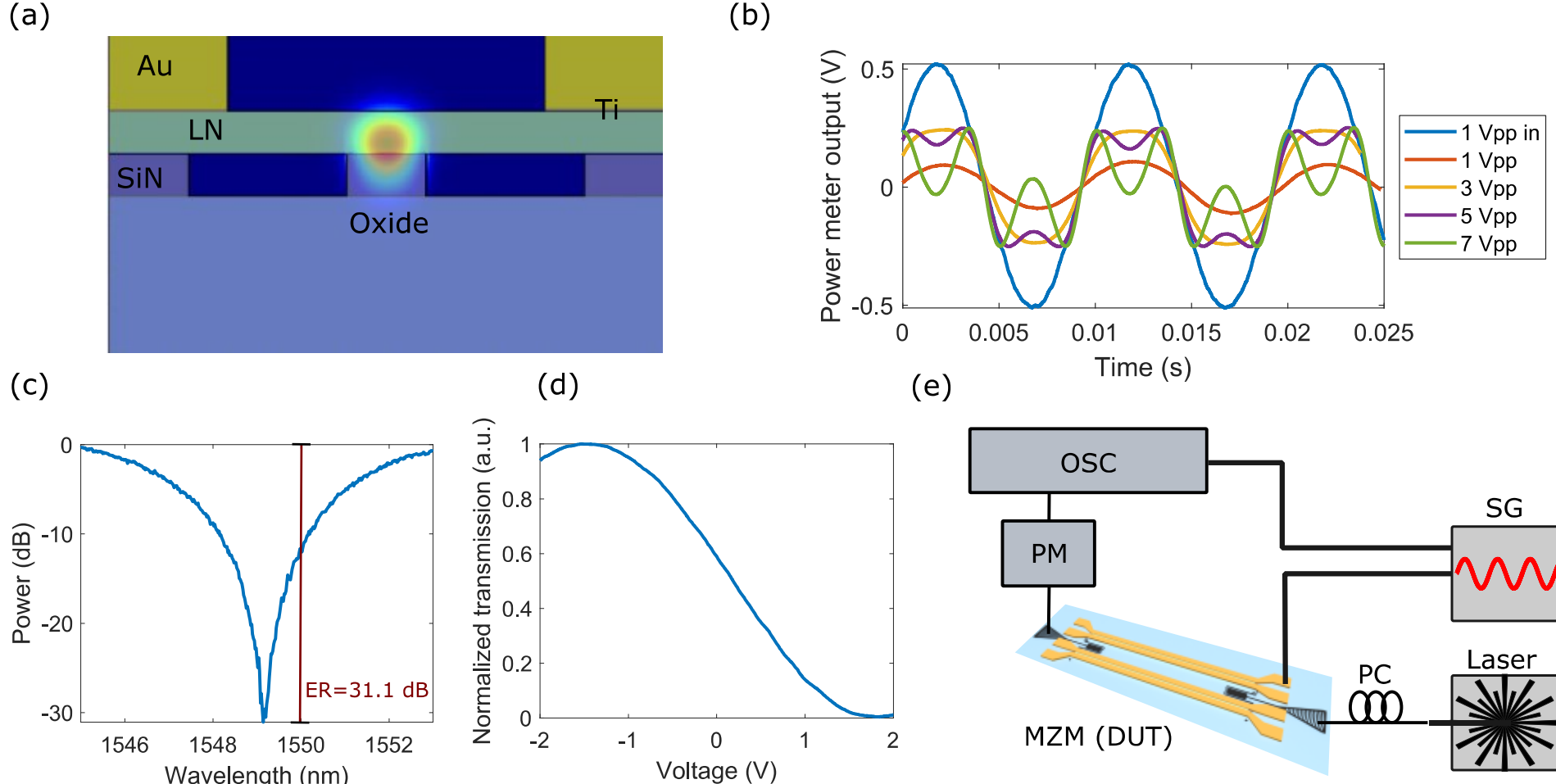

Fig. 7. MZM measurement data. (a) Mode cross-section simulation. (Au: 1 um, Ti: 15 nm, LN:300 nm, SiN: 300nm). (b) Example of the optical responses to sinusoidal electrical drive signals with different peak-to-peak voltage. (c) Extinction ratio (ER) extracted from an optical wavelength sweep. (d) Output power as a function of the applied voltage, from which the $V_\pi$ can be obtained. (e) Measurement setup which consists of a laser (optical input), a power meter (PM) to monitor the optical output, a signal generator (SG) to steer the electrical input and an oscilloscope (OSC) to monitor the electrical signals.

The measurements are carried out up to a frequency of 35 GHz and the de-embedded $|S_{21}|$ is plotted in Fig. 8.b. This $|S_{21}|$ parameter is the result after de-embedding the probe and the reference setup. After some resonances at low frequencies, a rather flat frequency response up to at least 35 GHz. The steep roll-off before 3 GHz can probably be attributed to an impedance mismatch of the electrodes and the probes, which is described in ref [25].

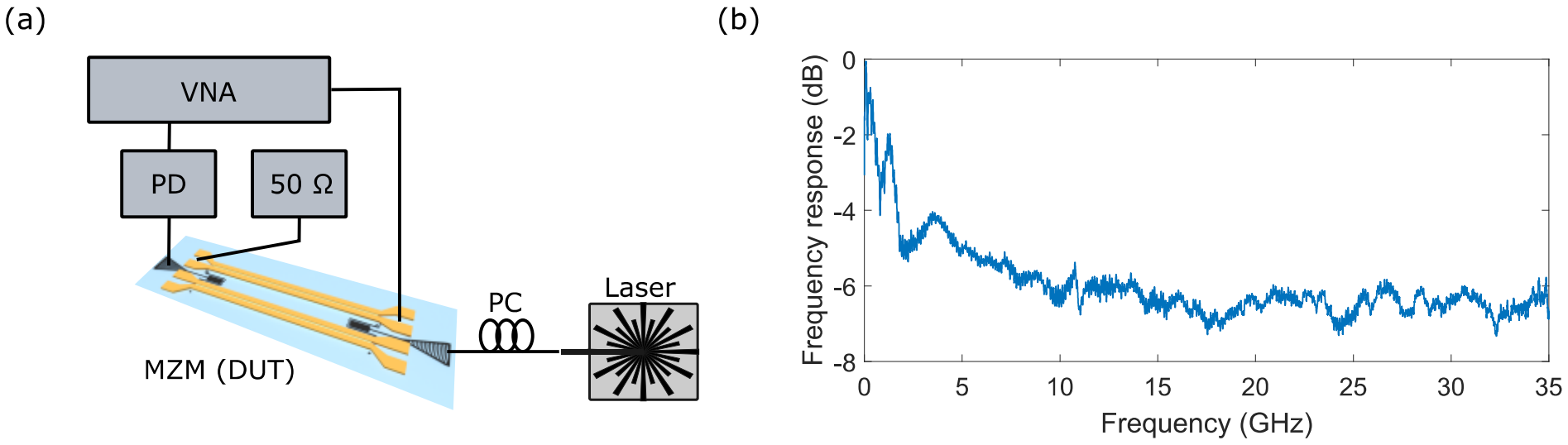

Fig. 8: (a) High speed measurement setup. VNA: vector network analyser, PD: photodiode, DUT: device under test, PC: polarization controller. (b) Frequency response of the modulator, obtained with the setup in (a).

Looking at other heterogeneous integration techniques in literature, LN-on-SiN modulators with a bandwidth of 30.6 GHz and 37 GHz were reported in [16] and [15] respectively. Neglecting the steep roll-off at the start, this is comparable to the bandwidth of this work. Yet, we believe that the modulator demonstrated in this work is not based on the most optimised design compared to the state of the art. For example, the electrode gap can be optimised, high impedance cladding and high resistivity substrate can be used or the overlap with the substrate can be minimised. However, it already effectively demonstrates the efficiency of integrating centimetre-scale LN devices through µTP. In future work, advanced designs using this versatile integration technique can be explored to further enhance the performance.

## 5. Discussion

We employed a back-end integration approach based on micro-transfer printing to ensure compatibility with mature Si-SiN platforms, enabling seamless co-integration with key components such as waveguides, splitters, grating couplers, heaters, and filters. As a proof of concept, we demonstrated the integration of TFLN onto a SiN platform fabricated in a short-loop run using e-beam lithography and RIE etching. However, applying this technique to a mature platform with lower propagation loss would further enhance performance while leveraging the available passive components of such platforms. The integration of TFLN introduces additional optical loss, reaching up to 1 dB/cm in this work. This increase is attributed to surface roughness resulting from encapsulation removal.

This technique also enables reversing the poling axis, as demonstrated in this work by printing the coupons in a push-pull configuration. This allows for complex electrode designs, such as ground-signal-signal-ground (GSSG), which are challenging to implement on monolithic LN platforms or via wafer bonding. Consequently, modulators can be driven by differential drivers, enhancing linearity by suppressing even-order harmonics and improving resilience to electromagnetic interference and power supply noise-critical aspects for multi-channel devices. Further studies are underway to explore these advantages in greater detail.

Additionally, micro-transfer printing ensures efficient utilisation of the costly TFLN material. For instance, a single 4-inch TFLN wafer can yield over 10,000 coupons. While wafer-scale heterogeneous integration using micro-transfer printing on 200 mm and 300 mm wafers remains a challenge, promising advancements in the field are paving the way for the widespread adoption of this technique in mass production.

All in all, heterogeneous integration of TFLN on silicon photonics via transfer printing presents another opportunity to enhance high-speed data communication systems, complementing previously demonstrated approaches such as LNOI and wafer-bonded TFLN platforms.

## 6. Conclusion

In this work, micro-transfer printing of up to 1 cm long coupons is demonstrated. They were printed on a patterned target to determine the losses and integrated into a push-pull Mach-Zehnder modulator to verify the electro-optical properties. The propagation loss is estimated to be around 0.9 dB/cm. Additionally, the insertion loss of the coupons is measured at 1.8 dB. With a $V_\pi$ of 3.2 V and modulation up to at least 35 GHz, the transfer printed cm-long TFLN in a push-pull modulator demonstrates a step towards the integration of LN on CMOS-compatible circuits, making it possible to have efficient modulators and nonlinear optical devices on mature platforms.

**Disclosures.** The authors declare no conflicts of interest.

**Data availability.** Data underlying the results presented in this paper are not publicly available at this time but may be obtained from the authors upon reasonable request.

**Funding.** The projects (40007560, 3G035722 and 101069447) have received funding from the FWO and F.R.S.-FNRS under the Excellence of Science programme and Horizon Europe.